# Two Effective Heuristics for Beam Angle Optimization in Radiation Therapy

**Hamed Yarmand, David Craft**

Department of Radiation Oncology

Massachusetts General Hospital and Harvard Medical School, Boston, MA 02114, USA

E-mail: yarmand.hamed@mgh.harvard.edu, dcraft@partners.org

**Corresponding author:**

Hamed Yarmand

Francis H. Burr Proton Therapy Center

Department of Radiation Oncology, MGH

Harvard Medical School

30 Fruit St., Boston, MA 02114, USA

Tell: +1-617-724-3665

Fax: +1-617-724-0368

## Abstract

**Purpose**: To improve the solution efficiency (i.e., reducing computation time while obtaining high-quality treatment plans) for the beam angle optimization problem (BAO) in radiation therapy treatment planning.

**Methods**: We formulate BAO as a mixed integer programming problem (MIP) whose solution gives the optimal beam orientation as well as optimal beam intensity map. We propose and investigate two novel heuristic approaches to reduce the computation time of the resultant MIP. One is a family of heuristic cuts based on the observation that the number of candidate beams is usually much larger than the number of beams used in the treatment plan, and therefore, it is less likely that "adjacent" beams are simultaneously used in the optimal treatment plan. The proposed cuts, referred to as "neighbor cuts", force the optimization system to choose one or a few beams from any set of adjacent beams. As a result, the search space and the computation time are reduced considerably. The second heuristic is a beam elimination scheme to eliminate beams with an insignificant contribution to deliver the dose to the tumor in the ideal plan in which all potential beams can be used simultaneously. Both heuristics can be added to any MIP formulation for BAO including the cases of coplanar/noncoplanar beams for intensity modulated radiation/proton therapy (IMRT and IMPT) and stereotactic body radiation therapy (SBRT). For the numerical experiments a clinical liver case (IMRT and SBRT) with 34 coplanar beams were considered.

**Results**: We first solved the corresponding MIP without the heuristics and recorded the optimal solution and the computation time. Then we incorporated the heuristics into the MIP and resolved it. Our results show that both heuristics reduce the computation time considerably while obtaining high-quality treatment plans.

**Conclusion**: This research incorporates two observations to improve the efficiency of the solution technique for BAO drastically: optimal beam configurations are typically a sparse set of well-spaced beams and optimal beams have a relatively large contribution to deliver the dose to the tumor in the ideal plan.

# 1 Introduction

Intensity modulation radiation therapy (IMRT) is one of the most successful external-beam radiation therapy delivery techniques for many sites of cancer due to its capability in delivery of highly complex dose distributions, hence delivering sufficient dose to the tumor and minimal dose to the surrounding organs at risk (OARs). Another delivery technique which is characterized by delivering a high amount of dose in a short period of time is stereotactic body radiation therapy (SBRT). Mathematical techniques are often exploited to obtain high-quality treatment plans. One important element of a treatment plan is the orientation of the beams used to deliver the radiation. The beam angle optimization problem (BAO) is known to be highly non-convex with many local minima[1]. Furthermore, the fluence map optimization problem (FMO), which finds the optimal intensity for each beamlet (in case of IMRT) or aperture (in case of SBRT) and is used to evaluate the treatment plan associated with a specific beam orientation, is a large-scale linear program (LP). Therefore finding the optimal solution to BAO requires an impractically long computation time, especially in case of IMRT. As a consequence, many heuristics have been developed to find "good" solutions to BAO (e.g., [2-7]). One heuristic which has sometimes been used is beam elimination, i.e., identifying and eliminating beams which are less likely to be among the optimal orientation. For example, Wang et al. [3,8] use a small number of beams (e.g., three) to explore the solution space to determine the most and least preferred directions. Then they eliminate the least preferred beams and search the remaining solution space using a fast gradient search algorithm.

The BAO has sometimes been modeled as a mixed integer program (MIP) (e.g., [9-11]). Since the resultant MIP is large-scale, heuristics have often been used to obtain good solutions in a reasonable amount of time. For example, Lim et al.[10] follow a beam elimination approach and use a scoring method to iteratively eliminate insignificant angles until a predetermined number of beams is reached. However, they admit that their iterative method do not work well with large number of candidate beams. Lim and Cao[11] use a branch and prune technique (based on a merit score function) combined with local search in a two-phase approach to find clinically acceptable solutions.

One way to reduce the computation time for a large-scale MIP is introducing constraints known as "valid inequalities". The purpose of adding these inequalities is to shrink the feasible region of the LP obtained by relaxing the integrality constraint so that it becomes closer to the convex hull of the feasible (integer) solutions, hence obtaining better bounds in the branch and bound (B&B) tree. This method, often combined with heuristics, has sometimes been used in radiation therapy treatment planning for reducing the computation time. For example, Gozbasi[12] derives valid inequalities for the MIP formulation of the volumetric-modulated arc therapy (VMAT) which decreases the solution time. Tuncel et al.[13] derive valid inequalities for FMO under dose-volume restrictions. Lee et al.[14] use disjunctive cuts for BAO along with other computational strategies including a constraint and column generation technique. Finally Taskin et al.[15-17] develop valid inequalities for the segmentation problem (i.e., the problem of converting the optimal fluence map to deliverable apertures).

The purpose of this paper is to introduce two novel heuristics which can be used alone or combined with other MIP-based BAO algorithms to obtain high-quality treatment plans in a reasonable amount of time. One is to add a set of heuristic inequalities, which we refer to as *neighbor cuts*, to the associated MIP in order to reduce the computation time. The idea is to exploit the intuition that the impacts of adjacent beams are similar, and therefore, it is less likely that adjacent beams are simultaneously chosen in the optimal beam orientation. The second heuristic is a beam elimination scheme in which beams with insignificant (dose) contribution in the ideal plan (in which all beams can be used without any restriction) are eliminated from consideration. Then the MIP is solved for the remaining candidate beams. Our numerical results for a clinical liver case we investigated show that both of these heuristics reduce the computation time considerably while attaining high-quality solutions for BAO.

## 2  Methods and Materials

### 2.1  Input data and dose calculation

The planning target volume (PTV) and OARs were identified using three-dimensional images (computed tomography (CT)). These volumetric data were then discretized into voxels for setting up the model instance which is a clinical liver case. Table 1 represents the list of structures for the model instance along with the number of voxels considered in each of the structures and the model parameters according to the clinical practice at Massachusetts General Hospital.

**Table 1.** Discretization of anatomical structures and model parameters.

| Structure  | Number of voxels | [Min,Max] dose (Gy) | Weight in Obj. Fun. |
|------------|------------------|---------------------|---------------------|
| PTV        | 5086             | [50,60]             | ---                 |
| Body       | 990              | [0,60]              | ---                 |
| Chest wall | 1893             | [0,60]              | 0.99                |
| Cord       | 42               | [0,45]              | 0.00                |
| Liver      | 743              | [0,60]              | 0.01                |

For the numerical experiments 34 uniformly distributed coplanar beams were considered. There were 34 beams with a gap of 83.72° between the 6$^{th}$ and 7$^{th}$ beams and a distance of 8.37° between two consecutive beams (see Figure 1). The methodology and solution technique presented in this paper can be directly applied to noncoplanar beams as well. Each beam was divided into 113-144 beamlets of size $1 \times 1\ cm^2$ (all beamlets go through PTV). This results in a large set of candidate beamlets for delivering the dose. For each beamlet, we used CERR (A Computational Environment for Radiotherapy Research) [18] to calculate the dose per monitor unit intensity to a voxel. The total dose per intensity deposited to a voxel is equal to the sum of dose per intensity deposited from each beamlet.

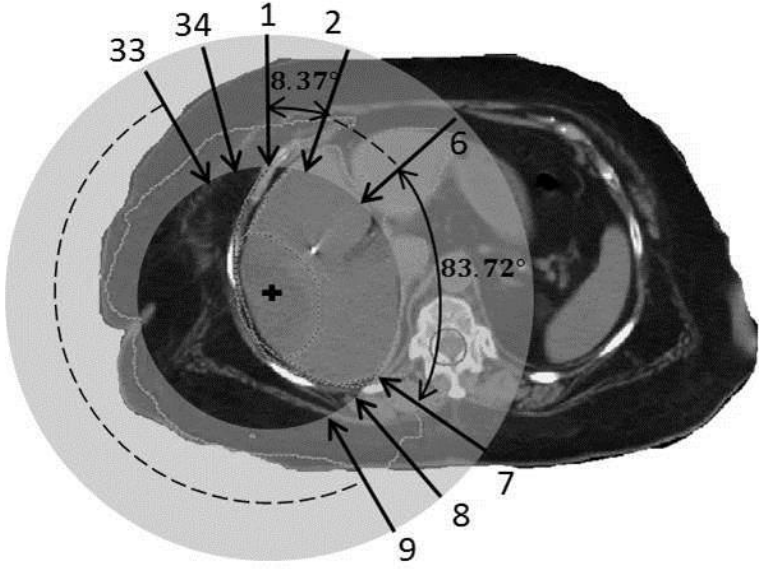

**Figure 1.** Position of 34 coplanar beams for the liver case.

## 2.2  Generating the pool of apertures for SBRT

In SBRT open fields or apertures that block a portion of the field are used to deliver the dose. Each aperture can be considered as a set of beamlets in a beam. Beamlets can be on/off in each beam to create apertures with different shapes. The only limitation is that in each row (corresponding to one leaf pair of the multileaf collimator) beamlets which are on should be located consecutively. To generate the pool of diverse candidate apertures first we calculated a *contribution score* for each beamlet based on its contribution to the dose delivery to PTV and OARs. Let $N$ denote the set of candidate beams, which could be coplanar or non-coplanar. Also let $L$ denote the set of OARs, respectively. For each beam $i \in N$, let $B_i$ denote the set of beamlets associated with beam $i$ and $D_{ijk}$ denote the dose per monitor unit intensity contribution to voxel $k$ from beamlet $j$ in beam $i$. The contribution score of beamlet $j$ in beam $i$, denoted by $\alpha_{ij}$, is calculated as follows.

$$\alpha_{ij} = \sum_{k \in V_p} D_{ijk} - \sum_{l \in L} \left( \lambda_l \sum_{k \in V_l} D_{ijk} \right) \quad , \quad i \in N, j \in B_i, \tag{1}$$

where $V_p$ and $V_l$ denote the set of voxels in PTV and in structure $l \in L$, respectively, and $\lambda_l \geq 0$ denotes the weight of structure $l \in L$ in calculating the contribution score. In the next step we employed a maximum subsequence sum algorithm to find the subsequence of beamlets in each row with the maximum subsequence sum of contribution scores. These subsequences formed the aperture with the maximum sum of beamlets' contribution scores.

As equation (1) demonstrates, the contribution scores, and hence the resultant pool of apertures, depend on $\lambda_l, l \in L$. If $\lambda_l > 0$, then the beamlets which deliver more dose to structure $l$ would have a lower contribution score, and hence would be less likely to be among the beamlets forming the resultant aperture. In other words, if $\lambda_l > 0$, then the resultant aperture avoids delivering a high level of dose to structure $l$. Accordingly we considered $|L| + 2$ different apertures for each beam: one aperture obtained by setting $\lambda_l = \bar{\lambda} > 0, l \in L$ (the smallest aperture which avoids all OARs), one aperture obtained by setting $\lambda_l = 0, l \in L$ (the largest aperture which is the beam's-eye-view of the target), and $|L|$ apertures obtained by setting $\lambda_{l'} = \bar{\lambda} > 0$ for $l' \in L$ and setting $\lambda_l = 0, l \in L, l \neq l'$ (this aperture only avoids structure $l'$). Therefore the pool of apertures contained $(|L| + 2)|N|$ candidate apertures.

Larger values of $\bar{\lambda}$ increase the importance of avoiding OARs, and hence, result in smaller apertures (except for the beam's-eye-view which remains unchanged). The converse is true for smaller values of $\bar{\lambda}$. In order to find the best value, we considered different values for $\bar{\lambda} \in [0.1, 1000]$. To evaluate the performance of the resultant pool of apertures for each value of $\bar{\lambda}$, we used the objective function as well as the dose distribution associated with the ideal plan in which all $|N|$ beams can be used (see Remark 2 in Section 2.3). We found that for the clinical liver case we investigated $\bar{\lambda} = 1$ resulted in the best treatment plan. Thus we generated the pool of apertures with $\bar{\lambda} = 1$.

## 2.3 Mixed integer programming treatment planning model

The basis of our model is a MIP which simultaneously optimizes the beam orientation as well as the beam intensity. In a treatment plan, each beamlet $j$ in each beam $i$ has a specific intensity $x_{ij} \geq 0$. If the clinically prescribed lower and upper bounds on the received dose at voxel $k$ are denoted by $C_k$ and $U_k$, respectively, then the basic dosimetric constraints can be represented as follows.

$$\sum_{i \in N} \sum_{j \in B_i} D_{ijk} x_{ij} \geq C_k \;, \; \forall k \quad \text{and} \quad \sum_{i \in N} \sum_{j \in B_i} D_{ijk} x_{ij} \leq U_k \;, \; \forall k, \qquad (2)$$

The lower and upper bounds on the received dose are usually the same for all voxels in a structure. The values we have used for these bounds for different structures are presented in Table 1.

Another restriction is the number of beams which can be used to deliver the radiation, which directly impacts the delivery time. Let $y_i$ be a binary variable whose value is 1 if beam $i$ is used and 0 otherwise. Also let $N_{max}$ be the maximum number of beams desired in the optimal plan. The following constraints limit the total number of beams used in the final plan and ensure that beamlet intensities are zero for unused beams.

$$\sum_{i \in N} y_i \leq N_{max} \quad \text{and} \quad x_{ij} \leq M_i y_i \quad, \quad i \in N, j \in B_i, \tag{3}$$

where $M_i$ is a positive constant and can be chosen based on the maximum possible intensity emitted from beam $i$. We have set $M_i = 35$ for all beams $i \in N$ based on the maximum possible beam intensity. The purpose of using the mathematical model is to minimize dose to OARs while keeping dose to the tumor above a predetermined value, which is presented by the lower bound constraint in (2) for voxels inside the tumor. Therefore the objective function in our model is to minimize the average dose to OARs. The average dose is calculated by taking average of dose over voxels in each OAR. We use average dose instead of absolute dose to eliminate the impact of different number of voxels in OARs. Let $L$ denote the set of OARs. Also let $w_l$ denote the importance weight of structure $l \in L$ (we can assume $\sum_{l \in L} w_l = 1$). The objective function of the MIP is

$$\min \sum_{l \in L} w_l (\sum_{i \in N} \sum_{j \in B_i} (D_{ij}^l) x_{ij}), \tag{4}$$

where $D_{ij}^l$ is the average dose per monitor unit intensity contribution to structure $l \in L$ from beamlet $j$ in beam $i$, i.e.,

$$D_{ij}^l = \frac{1}{|V_l|} (\sum_{k \in V_l} D_{ijk}), \tag{5}$$

where $V_l$ represents the set of voxels in structure $l \in L$. The importance weights $w_l, l \in L$ are determined by the planner based on his/her experience or, for example, by navigating through the ideal Pareto surface and choosing the desired point (e.g., see Craft *et al* [19]). For our numerical experiments we have used the values reported in Table 1 which were found by evaluating the dose distribution of the ideal plan. Note that we have considered a zero weight for cord because the dose to cord has been limited to 45 Gy as a constraint.

The objective function in (4) and the constraints in (2) and (3) form the MIP as the basis of our mathematical model. We emphasize that the neighbor cuts we discuss next can be added to any MIP formulation for BAO in radiation therapy treatment planning with any objective function or constraints. The key characteristic of the MIP formulation, which allows use of neighbor cuts, is representation of inclusion or exclusion of beams with binary variables.

**Remark 1**: In IMRT each beam consists of several beamlets while in SBRT each beam consists of several apertures. Therefore the problem formulation is the same if we replace "beamlets" for IMRT with "apertures" for SBRT. Thus we use the same MIP in (2)-(4) for BAO for SBRT with this difference that for SBRT we interpret $B(i)$ as the set of apertures associated with beam $i$, $D_{ijk}$ as the dose per monitor unit intensity contribution to voxel $k$ from aperture $j$ in beam $i$, and $x_{ij}$ as the intensity of aperture $j$ in beam $i$. For SBRT we calculate $D_{ijk}$ by summing the dose per monitor unit intensity contribution to voxel $k$ from all beamlets in aperture $j$ (in beam $i$). Note that in the optimal solution of SBRT more than one aperture with different intensities might be chosen from a beam.

**Remark 2:** The ideal plan is found by solving the associated LP which only includes continuous variables for intensity of each beamlet (in case of IMRT) or aperture (in case of SBRT). This LP is formed by the objective function in (4) and constraints in (2).

## 2.4 Neighbor cuts

The number of potential beams is often much larger than the number of beams which will actually be used. For example, assuming coplanar beams with a grid of 5°, there would be 36 potential beams. However, usually a few of them, e.g., 5 to 7 beams, are selected. Often the selected beams are relatively far apart from each other around the isocenter. In other words, it is less likely that adjacent beams, which have a similar impact, be selected simultaneously in the optimal orientation. This intuition is the basis of our proposed heuristic cuts: we add constraints, referred to as neighbor cuts, to the MIP which allow selection of at most one or a few of the beams in every set of adjacent beams (SAB), i.e., a set whose beams are pairwise adjacent. Therefore the general form of neighbor cuts is

$$\sum_{i \in \Omega_r} y_i \leq T_r \quad , \quad r = 1, 2, \dots, R, \tag{6}$$

where $\Omega_r, r = 1, 2 \dots, R$ represent $R$ different SABs and $T_r$ denotes the maximum number of adjacent beams which can be selected from SAB $r$ (note that $T_r < |\Omega_r|$ otherwise constraints (6) will be redundant). The definition of *adjacency* is based on the distance (in degrees) between different beam angles. For example, one might define two adjacent beams as two beams with a distance of at most 10°. In case of uniform distribution of coplanar candidate beams, which is often considered in clinical practice, the adjacency can be defined based on the order in which beams are located around the isocenter. For example, if beams are numbered from 1 to 36, two beams might be defined to be adjacent if their order differs at most by two (it means that every three consecutive beams, e.g., beams 7, 8, and 9, form a SAB). In this case all SABs include the same number of candidate beams, and hence, the same maximum allowed number of beams can be considered for all SABs (i.e., $T_r = T, r = 1, 2, \dots, R$). Therefore in case of uniform and coplanar distribution of candidate beams the neighbor cuts can be represented as follows.

$$\sum_{s=0}^{S-1} y_{i+s} \leq T \quad , \quad i \in N, \tag{7}$$

where $S \geq 2$ denotes the *neighborhood size* defined as the maximum difference in the order of two adjacent beams. Since the beams are repeated after the $N$th beam, we define the index $i + s \equiv i + s - |N|$ if $i + s > |N|$. Also in order to use maximum number of beams, $N_{max}$, in the final plan we must have $N_{max} \leq \left\lfloor \frac{|N|}{S} \right\rfloor T$, where $\lfloor a \rfloor$ returns the largest integer which is smaller than or equal to $a$. We assume coplanar beams with a uniform distribution in the rest of this paper and for the clinical case we investigate. However, the definition of adjacent beams, and hence the application of neighbor cuts, can be easily extended to non-coplanar beams in the three-dimensional space.

Two parameters should be determined in order to use the neighbor cuts. One is the neighborhood size, $S$, and the other is the maximum allowed number of beams in each SAB, $T$. A larger value for $S$ or a smaller value for $T$ imposes tighter restrictions on the original MIP as a result of adding constraints (7). Therefore, in general, it reduces the quality of the heuristic solution and also reduces the computation time further. The converse is in general true for a smaller value for $S$ or a larger value for $T$. It is desired to use the value of $S$ which is large enough to reduce the computation time considerably but small enough to avoid infeasibility and low-quality solutions.

## 2.5 Beam elimination

In a treatment plan, some beams contribute more and some contribute less to delivery of radiation to the tumor. It is quite likely that beams with a lower contribution will not be used if the maximum number of beams is reduced. This is the basis of our beam elimination scheme. First we find the ideal plan by solving the associated LP. In the next step we calculate the percentage contribution of each beam to the dose delivered to the tumor in the ideal plan, which we refer to as *dose contribution*. Then we eliminate the beams with lower dose contributions and solve the BAO for the remaining beams. Similar to neighbor cuts, the beam elimination can be applied to both coplanar and noncoplanar beams.

One approach for beam elimination is to determine a dose contribution threshold, denoted by $\varepsilon$, as the elimination criterion. In this approach all beams with a dose contribution lower than $\varepsilon$ are eliminated. This approach provides a control (in fact a lower bound) on the dose contribution of the remaining beams. We have followed this approach in this research. Another approach is to determine the number of beams we want to remain in the pool of candidate beams. In this approach all candidate beams are ranked according to their dose contribution and then a specific number of beams with the least dose contributions are eliminated. This approach provides a means to control the computation time to some extent because the computation time (i.e., the time required to solve the MIP) directly depends on the number of candidate beams. However, this approach might not be appropriate if the dose contributions of different beams are close to each other. In practice, a combination of these two approaches might be used based on the planner expertise and the realized dose contributions in the ideal plan. In particular, note that these two approaches result in the same set of remaining beams with appropriate parameters.

For smaller numbers of eliminated beams, the size of the associated MIP would be larger, and the computation time would be longer. However, the resultant heuristic solution would, in general, have a higher quality since the optimal beams have been chosen from a larger pool of candidate beams. The converse is in general true for larger numbers of eliminated beams.

## 2.6 Computation

To solve the MIP we used CPLEX (version 12.5) on a PC with 2 Intel Xeon X5650 (2.66 GHz) CPU and 48 GB of RAM.

## 3 Results and Discussion

The results for the IMRT and SBRT of the liver case are presented in Tables 2 and 3, respectively. We used the LP with the objective function in (4) and constraints in (2) to find the ideal plan. These plans for the IMRT and SBRT of the liver case are represented in the first row in Tables 2 and 3. We used the MIP in (2)-(4) to solve the BAO (for IMRT and SBRT) for the clinical liver case reported in Table 1. First we solved the original MIP for $N_{max} = 5,7,9$ and found the optimal solution (bold rows in Tables 2 and 3). Then we incorporated the neighbor cuts and the beam elimination heuristics into the MIP and resolved it. While 33 beams were used in the IMRT ideal plan, only 13 beams were used in the SBRT ideal plan. Therefore we eliminated the unused beams for beam elimination in the SBRT case. However, for IMRT, we considered two different values for the dose contribution threshold, $\varepsilon = 2\%$ and $\varepsilon = 5\%$, which resulted in elimination of 20 (beams 8, 9,…, 26, 28, 29) and 26 beams (beams 5, 6,…, 30, 33), respectively.

In Tables 2 and 3, "optimality gap" and "time reduction" are calculated compared to the corresponding MIP with no heuristics (i.e., compared to the corresponding bold rows). Also in Table 3 the underlined and double underlined beams represent beams with two and three apertures in the optimal solution, respectively.

Table 2. Numerical results for the IMRT of the liver case.

| $N_{max}$ | $\varepsilon$ (%) | S | T | Num. Beams | Obj. value | Opt. gap (%) | Comp. time (h) | Time Reduc. (%) | Beams used |
|---|---|---|---|---|---|---|---|---|---|
| 34 | --- | --- | --- | 33 | 7.63 | 0.00 | 0.01 | --- | 1,2,…,12,14,15,…,34 |
| **9** | --- | --- | --- | **9** | **8.42** | **0.00** | **26.66** | --- | **1,2,3,4,5,7,31,32,34** |
| 9 | --- | 2 | 1 | 9 | 8.66 | 2.76 | 3.54 | 86.7 | 2,4,6,7,26,28,30,32,34 |
| 9 | --- | 3 | 1 | 9 | 9.01 | 6.99 | 3.52 | 86.8 | 1,4,7,10,13,21,25,28,32 |
| 9 | --- | 3 | 2 | 9 | 8.51 | 1.01 | 20.21 | 24.2 | 1,2,4,6,7,26,28,31,32 |
| 9 | 2 | --- | --- | 9 | 8.42 | 0.00 | 0.36 | 98.7 | 1,2,3,4,5,7,31,32,34 |
| 9 | 5 | --- | --- | 7 | 8.95 | 6.27 | 0.01 | 99.98 | 1,2,3,4,31,32,34 |
| **7** | --- | --- | --- | **7** | **8.78** | **0.00** | **32.07** | --- | **1,2,3,4,5,32,34** |
| 7 | --- | 2 | 1 | 7 | 8.92 | 1.56 | 8.99 | 72.0 | 2,4,6,7,26,32,34 |
| 7 | --- | 3 | 1 | 7 | 9.16 | 4.24 | 4.73 | 85.3 | 1,4,7,10,25,28,32 |
| 7 | --- | 3 | 2 | 7 | 8.81 | 0.26 | 24.62 | 23.2 | 1,2,4,6,7,32,33 |
| 7 | --- | 4 | 1 | 7 | 9.26 | 5.36 | 2.26 | 93.0 | 1,5,7,11,21,26,30 |
| 7 | --- | 4 | 2 | 7 | 8.90 | 1.34 | 22.95 | 28.4 | 1,2,5,6,,7,31,32 |
| 7 | 2 | --- | --- | 7 | 8.78 | 0.00 | 0.72 | 97.7 | 1,2,3,4,5,32,34 |
| 7 | 5 | --- | --- | 7 | 8.95 | 1.90 | 0.01 | 99.98 | 1,2,3,4,31,32,34 |
| **5** | --- | --- | --- | **5** | **9.29** | **0.00** | **20.51** | --- | **1,3,5,32,34** |
| 5 | --- | 2 | 1 | 5 | 9.30 | 0.10 | 5.46 | 73.4 | 1,3,5,7,33 |
| 5 | --- | 3 | 1 | 5 | 9.48 | 2.06 | 4.16 | 79.7 | 1,4,7,26,32 |
| 5 | --- | 3 | 2 | 5 | 9.29 | 0.00 | 9.43 | 54.0 | 1,3,5,32,34 |
| 5 | --- | 4 | 1 | 5 | 9.58 | 3.16 | 3.06 | 85.1 | 2,6,7,26,32 |
| 5 | --- | 4 | 2 | 5 | 9.30 | 0.10 | 9.17 | 55.3 | 1,3,5,7,33 |
| 5 | 2 | --- | --- | 5 | 9.29 | 0.00 | 0.78 | 96.2 | 1,3,5,32,34 |
| 5 | 5 | --- | --- | 5 | 9.40 | 1.17 | 0.06 | 99.7 | 1,3,4,32,34 |

To illustrate the quality of the resultant heuristic plans, we have compared the dose-volume histogram (DVH) for the optimal plan for $N_{max} = 7$ with the generated plan with $S = 3$ and $T = 1$ in Figure 2. Note that although the cord dose has gone up substantially, it is still well below the constraint of 45 Gy.

As Tables 2 and 3 show, the neighbor cuts have a good performance in reducing the computation time, especially for larger values of $S$ and smaller values of $T$. They reduce the computation time considerably while keeping the optimality gap small. In some cases, especially in case of SBRT, adding the neighbor cuts results in finding the optimal beam orientations (i.e., a zero optimality gap). Note that the optimality gap resulted from adding the neighbor cuts is in general smaller for smaller values of $N_{max}$ due to sparser distribution of beams. The reduction in computation time is more substantial for IMRT although it is accompanied with a larger optimality gap compared to SBRT.

**Table 3**. Numerical results for the SBRT of the liver case.

| $N_{max}$ | Beam Elim. | S | T | Num. Beams | Obj. value | Opt. gap (%) | Comp. time (s) | Time Reduc. (%) | Beams used |
|---|---|---|---|---|---|---|---|---|---|
| 34 | --- | --- | --- | 13 | 14.21 | 0.00 | 12.3 | --- | 1,2,3,5,6,7,12,13,23,26,27,33,34 |
| **9** | --- | --- | --- | **9** | **14.23** | **0.00** | **93.4** | --- | **1,3,6,7,12,13,23,26,33** |
| 9 | --- | 2 | 1 | 9 | 14.25 | 0.15 | 38.7 | 58.6 | 1,3,6,7,12,13,23,26,33 |
| 9 | --- | 3 | 1 | 9 | 14.49 | 1.80 | 52.9 | 43.3 | 1,5,7,13,26,32 |
| 9 | --- | 3 | 2 | 9 | 14.23 | 0.00 | 33.3 | 64.3 | 1,3,6,7,12,13,23,26,33 |
| 9 | Yes | --- | --- | 9 | 14.23 | 0.00 | 50.1 | 46.3 | 1,3,6,7,12,13,23,26,33 |
| **7** | --- | --- | --- | **7** | **14.26** | **0.00** | **125.8** | --- | **1,3,6,7,13,26,33** |
| 7 | --- | 2 | 1 | 7 | 14.26 | 0.00 | 86.4 | 31.4 | 1,3,6,7,13,26,33 |
| 7 | --- | 3 | 1 | 6 | 14.49 | 1.59 | 51.7 | 58.9 | 1,5,7,13,26,32 |
| 7 | --- | 3 | 2 | 7 | 14.26 | 0.00 | 102.2 | 18.8 | 1,3,6,7,13,26,33 |
| 7 | --- | 4 | 1 | 5 | 14.50 | 1.66 | 41.1 | 67.3 | 1,5,7,13,26 |
| 7 | --- | 4 | 2 | 7 | 14.26 | 0.00 | 79.8 | 36.5 | 1,3,6,7,13,26,33 |
| 7 | Yes | --- | --- | 7 | 14.26 | 0.00 | 61.9 | 50.8 | 1,3,6,7,13,26,33 |
| **5** | --- | --- | --- | **5** | **14.47** | **0.00** | **328.8** | --- | **1,3,6,7,12** |
| 5 | --- | 2 | 1 | 5 | 14.47 | 0.00 | 227.9 | 30.7 | 1,3,6,7,12 |
| 5 | --- | 3 | 1 | 5 | 14.50 | 0.19 | 49.8 | 84.8 | 1,5,7,13,26 |
| 5 | --- | 3 | 2 | 5 | 14.47 | 0.00 | 202.7 | 38.3 | 1,3,6,7,12 |
| 5 | --- | 4 | 1 | 5 | 14.50 | 0.19 | 38.7 | 88.2 | 1,5,7,13,26 |
| 5 | --- | 4 | 2 | 5 | 14.47 | 0.00 | 170.9 | 48.0 | 1,3,6,7,12 |
| 5 | Yes | --- | --- | 5 | 14.47 | 0.00 | 98.1 | 70.2 | 1,3,6,7,12 |

The beam elimination heuristic also has a very good performance in reducing the computation time, especially for IMRT. In particular, in all SBRT cases and also IMRT cases with $\varepsilon = 2\%$ the beam elimination results in finding the optimal beam orientation (i.e., a zero optimality gap).

We observe that in some cases the number of beams in the optimal orientation is smaller than $N_{max}$. As expected, this is more common for SBRT since arbitrary modulation, which makes all beams attractive, is not possible with only a few predefined apertures available. Another observation is that reducing the number of beams does not have a substantial impact on the plan quality for SBRT compared to IMRT (compare the objective values of the bold rows).

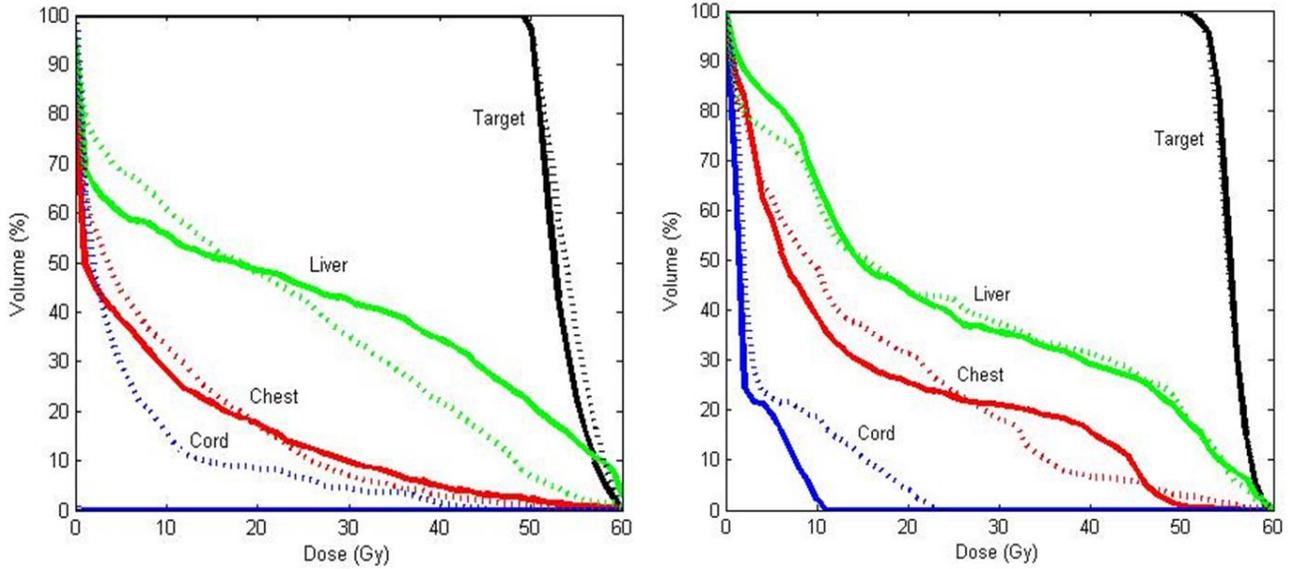

**Figure 2.** DVH for the optimal plan for $N_{max}$ = 7 (solid lines) and the generated plan with
$S$ = 3 and $T$ = 1 (dotted lines) for IMRT (left) and SBRT (right).

## 4 Conclusion

We proposed two heuristics for reducing the computation time for BAO which can be applied to any MIP-based formulation, e.g., IMRT, IMPT, and SBRT. One is adding the neighbor cuts to the associated MIP based on the intuition that the impact of adjacent beams are similar and it is less likely that two adjacent beams are used in the optimal orientation. The other is a beam elimination scheme in which beams with insignificant (dose) contribution in the ideal plan are eliminated from consideration. These heuristics can be applied alone or combined with other heuristics to find high-quality treatment plans in a reasonable amount of time. Our numerical results for IMRT and SBRT of a clinical liver case showed that both heuristics were capable of reducing the computation time considerably while attaining high-quality solutions, hence accelerating the treatment planning process.

## Acknowledgments

This research was supported by the Federal Share of program income earned by Massachusetts General Hospital on C06 CA059267, Proton Therapy Research and Treatment Center, and by RaySearch Laboratories.